\documentclass[pra,twocolumn,showpacs,amsmath,amssymb,superscriptaddress]{revtex4}
\usepackage{graphicx}
\usepackage{bm}
\usepackage{amsmath,amssymb}
\begin{document}

\title{Visualization of superposition of macroscopically distinct states}

\author{Tomoyuki Morimae}
\email{morimae@ASone.c.u-tokyo.ac.jp}
\affiliation{
Department of Basic Science, University of Tokyo, 
3-8-1 Komaba, Tokyo 153-8902, Japan
}
\affiliation{
PRESTO, Japan Science and Technology Corporation,
4-1-8 Honcho, Kawaguchi, Saitama, Japan
}
\affiliation{
Japan Society for the Promotion of Science, 8 Ichibancho, Tokyo 102-8471, Japan}
\author{Akira Shimizu}
\email{shmz@ASone.c.u-tokyo.ac.jp}
\affiliation{
Department of Basic Science, University of Tokyo, 
3-8-1 Komaba, Tokyo 153-8902, Japan
}
\affiliation{
PRESTO, Japan Science and Technology Corporation,
4-1-8 Honcho, Kawaguchi, Saitama, Japan
}
\date{\today}
            
\begin{abstract}
We propose 
a method of visualizing 
superpositions of macroscopically distinct states in
many-body pure states.
We introduce a visualization function, 
which is a 
coarse-grained quasi joint probability density for two or more hermitian additive
operators.
If a state contains superpositions of macroscopically distinct states,
one can visualize them
by plotting the visualization function 
for
appropriately taken operators.
We also explain how to efficiently find 
appropriate operators for a given state.
As examples,
we visualize 
four states containing superpositions of macroscopically distinct states: 
the ground state of the XY model,
that of the Heisenberg antiferromagnet,
a state  
in Shor's factoring algorithm,
and a state in Grover's quantum search algorithm. 
Although the visualization function can take negative values, 
it becomes non-negative 
(hence becomes a 
coarse-grained joint probability density)
if the characteristic width of the coarse-graining function
used in the visualization function is sufficiently large.

\end{abstract}
%------------------------------------------------------------
\maketitle  

\section{introduction}

Visualization functions,
such as the Wigner distribution function~\cite{Wigner}
and the Husimi function~\cite{husimi},
are very useful.
By plotting them,
one can visualize quantum states
to understand 
structure of these states. 
Furthermore, because these functions are, in some senses,
probability densities, 
one can interpret various 
experimental results
by using these functions~\cite{Mandel}.
Although there are many methods of visualizing 
quantum states 
with small degrees of freedom~\cite{Mandel},
those of visualizing quantum many-body states 
are 
very few
~\cite{Wooters,Leonhardt,Hannay,Rivas}.
It is therefore important to develop
methods of visualizing quantum many-body states.

In quantum many-body systems, 
which include 
quantum computers~\cite{Nielsen,Grover,Shor,Ekert},
there are many states 
which contain
superpositions of macroscopically distinct 
states~\cite{Schrodinger,Leggett1,Nakamura,
Friedman,Wal,Mermin,Wakita,Grib,Cirac,SM,Morimae,SM05,Ukena,Ukena2,Boson}.
Existence of a superposition of macroscopically distinct states in  
a many-body pure state can be identified by an index 
$p$ $(1\le p\le 2)$~\cite{SM,Sugita,Morimae,Ukena,Ukena2}:
If a given pure state has $p=2$, it contains a superposition of 
macroscopically distinct states~\cite{SM,Morimae}.

If every macroscopic superposition 
could be reduced to
an equal-weight superposition of 
two macroscopically distinct states,
such as $\frac{1}{\sqrt{2}}
|0\cdots0\rangle
+\frac{1}{\sqrt{2}}|1\cdots1\rangle$,
visualization of macroscopic 
superpositions would be a trivial task.
However, there are 
many other states in which 
many macroscopically distinct states are superposed with
various weights~\cite{SM,Morimae,SM05,Ukena,Ukena2,Boson}.
Therefore it is also important to develop 
good methods of visualizing 
such complicated superpositions.

In this paper, we propose
a method of visualizing superpositions of macroscopically distinct states contained in
states
having $p=2$.
We first introduce a
function $\Xi(A_1,\cdots,A_m)$,
which is interpreted as a
coarse-grained quasi joint probability density
for 
hermitian additive operators $\hat{A}_1,\cdots,\hat{A}_m$.
We next explain how to find  
appropriate $\hat{A}_1,\cdots,\hat{A}_m$ efficiently
for a given pure state.
One can visualize superpositions of macroscopically distinct states
contained in a given pure state having $p=2$
by plotting $\Xi(A_1,\cdots,A_m)$
for appropriate $\hat{A}_1,\cdots,\hat{A}_m$.
As examples, we visualize four states  
having $p=2$: 
the ground state of the XY model,
that of the Heisenberg antiferromagnet,
a state in Shor's factoring algorithm~\cite{Shor},
and 
a state in Grover's quantum search algorithm~\cite{Grover}.
Although $\Xi$ can take negative values,
like the Wigner distribution function,
it becomes non-negative, hence becomes 
a coarse-grained joint probability
density, if the characteristic width of the coarse-graining
function used in $\Xi$ is sufficiently large.

This paper is organized as follows.
After briefly reviewing the index $p$ in the next section,
we introduce $\Xi$ in Sec.~\ref{cgQJPD}, and
explain how to find  
appropriate operators efficiently in Sec.~\ref{finding}.
We visualize 
four states in Sec.~\ref{examples}.
Discussion  
is given in Sec.~\ref{discussion}.

%----------------------------------------------------------
\section{\label{p}Index $p$}

To establish notation,
and for the convenience of the reader, 
we briefly review 
the index $p$ in this section.
For details, see Refs. 
\cite{SM,Ukena,Ukena2,Sugita,Morimae,SM05}.

We first fix the energy range of interest.
It determines
the degrees of freedom of an effective theory which describes the system 
under consideration.
We assume that the system is, in that energy range, 
described as an $N$-site lattice.
Throughout this paper, we assume that $N$ is large but finite.

For simplicity, we here consider only pure states, 
although the definition of superposition of
macroscopically distinct states 
has been successfully generalized to mixed states \cite{SM05}.
Furthermore, we assume that 
states are 
spatially homogeneous, or effectively homogeneous
as in quantum chaotic systems \cite{Sugita} or in quantum computers
\cite{Ukena,Ukena2}.
For such states, 
we can consider a family of similar states $\{|\psi^N\rangle\}_{N}$.
For example, %if we consider the index $p$ for 
each member of the family $\{ |E_0^N\rangle \}_N$ of the ground states 
of the XY model is 
the ground state $|E_0^N\rangle$ of the XY Hamiltonian
of an $N$-site system.

The index $p$ is defined for 
such families of similar states.
For simplicity, we represent a family of states $\{|\psi^N\rangle\}_{N}$
by a representative state $|\psi\rangle$ $(=|\psi^N\rangle)$.
The index $p$ $(1\leq p \leq 2)$ of $|\psi\rangle$
is then defined by
\begin{eqnarray}
\max_{\hat{A}}\langle\psi|(\Delta \hat{A})^2|\psi\rangle
=\mathcal{O}(N^p),
\label{defp}
\end{eqnarray}
where 
$\Delta \hat{A}\equiv
\hat{A}-\langle\psi|\hat{A}|\psi\rangle$ \cite{O},
and the maximum is taken over all hermitian additive operators $\hat{A}$.
Here, an additive operator $\hat{A}$ is a sum of local 
operators:
$\hat{A}=\sum_{l=1}^N\hat{a}(l)$,
where $\hat{a}(l)$ is a local operator, which is independent of $N$, on site $l$.
We do not assume that 
$\hat{a}(l')$ ($l' \neq l$) is the spatial translation of $\hat{a}(l)$.

If $p=2$, there is a hermitian additive operator which 
`fluctuates macroscopically'
in the sense that
the relative fluctuation 
does not vanish in the limit of $N\to\infty$:
\begin{eqnarray}
\frac{\sqrt{\langle\psi|(\Delta \hat{A})^2|\psi\rangle}}{N}
%=\frac{\sqrt{\mathcal{O}(N)}}{N}
\nrightarrow0\ \ (N\to\infty).
\label{relativefluctuation}
\end{eqnarray}
Because $|\psi\rangle$ is pure, 
the reason for the macroscopic fluctuation
is that
eigenstates of $\hat{A}$ corresponding to 
macroscopically distinct 
eigenvalues are superposed 
with sufficiently large weights in $|\psi\rangle$.
Here, two eigenvalues $A$ and $A'$ are macroscopically distinct
if and only if $A-A'=\mathcal{O}(N)$.
Therefore a pure state having $p=2$ contains 
a superposition of macroscopically distinct states
(see Refs.~\cite{Morimae,SM05} for detailed discussion).
On the other hand, if $p <2$,  
all additive operators `have
macroscopically definite values' 
in the sense that
relative fluctuations of all additive operators 
vanish as $N\to\infty$.
In this case, there is no superposition of macroscopically distinct states
in $|\psi\rangle$.
In short, one can judge whether a pure state 
contains a superposition of macroscopically distinct states or not
by calculating the index $p$.

There is an efficient method of calculating $p$.
For simplicity, we henceforth assume that 
each site of the lattice is a spin-$1/2$ system. 
For a given pure state $|\psi\rangle$,
we define the variance-covariance matrix (VCM) by
\begin{eqnarray}
V_{\alpha l, \beta l'} \equiv 
\langle\psi|\Delta
\hat{\sigma}_{\alpha}(l)\Delta
\hat{\sigma}_{\beta}(l')|\psi\rangle,
\label{VCM}
\end{eqnarray}
where 
$\alpha,\beta=x,y,z$;
$l,l'=1,2,\cdots,N$; 
$\hat{\sigma}_x(l)$, $\hat{\sigma}_y(l)$, and $\hat{\sigma}_z(l)$
are Pauli operators on site $l$.
The VCM is 
a $3N\times 3N$ hermitian non-negative matrix. 
If $e_1$ is the maximum eigenvalue 
of the VCM, then 
$e_1 = \mathcal{O}(N^{p-1})$, 
as shown in Appendix \ref{app:e1}.
One therefore has only to evaluate $e_1$
to calculate $p$.

%-----------------------------------------------------
\section{Visualization method}\label{visualize}

By calculating the index $p$,
one can 
judge whether a pure state
contains a superposition of macroscopically distinct states or not. 
From $p$ only, however, one cannot know 
detailed structures of the superposition of macroscopically distinct states,
including
which macroscopically distinct 
states are superposed 
and with what weights they are superposed.
In this section, we propose a method of
visualizing these structures of 
superpositions of macroscopically distinct states.

%-----------------------------
\subsection{Visualization function $\Xi$}
\label{cgQJPD}

Let $\hat{A}=\sum_l\hat{a}(l)$ and $\hat{B}=\sum_l\hat{b}(l)$
be hermitian 
additive operators.
We assume that $[\hat{A}, \hat{B}] \neq 0$, 
so that the joint probability distribution 
for $\hat{A}$ and $\hat{B}$ does not exist in general.

For macroscopic systems, one is usually interested in 
states in which typical values of  
additive operators are $\mathcal{O}(N)$ \cite{O}.
Typical values of 
$\hat{A}/N$ and $\hat{B}/N$ are therefore $\mathcal{O}(N^0)$.
Their commutator is small in the sense that
\begin{equation}
\left\| \left[ \frac{\hat{A}}{N}~,~\frac{\hat{B}}{N} \right] \right\|
= O \left(\frac{1}{N}\right),
\label{commute1}
\end{equation}
because 
$[\hat{a}(l),\hat{b}(l')]=0$ for $l\neq l'$~\cite{O,SM05,Grib}.
Since $N$ is large but finite, the above commutator does not vanish.
In real experiments, however, 
resolutions of measurements are limited.
Equation~(\ref{commute1}) therefore indicates that 
noncommutativity of additive operators
could not be detected 
for large $N$. 
This suggests that we may be able to introduce
a function which can be well regarded as 
a coarse-grained
joint probability density
for $\hat{A}$ and $\hat{B}$.
Note that the finite resolution is essential,
because 
noncommutativity, however small it is,
can be detected if the resolutions of experiments 
are high enough~\cite{xp}. 

We formulate the above idea as follows.
Consider the spectral decompositions of 
$\hat{A}$ and $\hat{B}$: 
\begin{equation}
\hat{A} = \sum_{A \in \mathbb{E}_{\hat A}} A \mathcal{P}_{\hat A}(A),\ 
\hat{B} = \sum_{B \in \mathbb{E}_{\hat B}} B \mathcal{P}_{\hat B}(B),
\end{equation}
where $\mathbb{E}_{\hat A}$ and $\mathbb{E}_{\hat B}$ 
are the spectra of $\hat{A}$ and $\hat{B}$, respectively, and 
$\mathcal{P}_{\hat A}(A)$ and 
$\mathcal{P}_{\hat B}(B)$ are 
the projection operators onto 
the eigenspaces of eigenvalues $A$ and $B$, respectively.
To take account of finite resolutions of experiments, 
we smear the projection operators to obtain 
\begin{eqnarray}
\overline{\mathcal{P}}_{\hat A}(A)
\equiv
\sum_{A' \in \mathbb{E}_{\hat A}} w_{\hat A}(A,A') 
\mathcal{P}_{\hat A}(A'),
\end{eqnarray}
and similarly for $\overline{\mathcal{P}}_{\hat B}(B)$.
Here, $A$ is a real continuous variable ($A\in\mathbb{R}$),
and $w_{\hat A}(A, A')$ 
is a coarse-graining function.
It centers at $A=A'$ with a characteristic width $W_{\hat A}$, 
and satisfies
\begin{eqnarray}
&&
w_{\hat A}(A, A') \geq 0
\quad \mbox{for all } A, A',
\\
&&
\int_{-\infty}^{+\infty}
w_{\hat A}(A, A') dA=1
\quad \mbox{for all } A'.
\end{eqnarray}
The coarse-graining functions $w_{\hat A}, w_{\hat B}$ 
should not have complicated forms; 
they should be physically reasonable ones. 
To be definite, 
we henceforth assume that $W_{\hat A}=W_{\hat B}=W$ and 
$w_{\hat A}(X,X')=w_{\hat B}(X,X')=w(X-X')$, where
\begin{equation}
w(X)={1 \over \sqrt{2 \pi} W}
\exp\left(-{X^2 \over 2W^2} \right).
\label{eq:w}\end{equation}

Clearly, 
$\overline{\mathcal{P}}_{\hat A}(A)$
and 
$\overline{\mathcal{P}}_{\hat B}(B)$
are non-negative hermitian operators satisfying 
\begin{equation}
\int_{-\infty}^{+\infty}
\overline{\mathcal{P}}_{\hat A}(A) dA
= 
\int_{-\infty}^{+\infty}
\overline{\mathcal{P}}_{\hat B}(B) dB
= \hat 1.
\end{equation}
They give coarse-grained 
probability densities 
$\Xi_{\hat A}$
and
$\Xi_{\hat B}$
for 
$\hat{A}$ and $\hat{B}$, respectively, for 
a given pure state $| \psi \rangle$ by
\begin{eqnarray}
&& 
\Xi_{\hat A}(A) =
\langle\psi|\overline{\mathcal{P}}_{\hat A}(A) |\psi\rangle
\quad (A \in \mathbb{R}),
\\
&&
\Xi_{\hat B}(B) =
\langle\psi|\overline{\mathcal{P}}_{\hat B}(B) |\psi\rangle
\quad (B \in \mathbb{R}).
\end{eqnarray}
Now we define 
\begin{equation}
\Xi(A,B) \equiv
\frac{1}{2}\langle\psi|
\overline{\mathcal{P}}_{\hat A}(A) 
\overline{\mathcal{P}}_{\hat B}(B)
+
\overline{\mathcal{P}}_{\hat B}(B)
\overline{\mathcal{P}}_{\hat A}(A) 
|\psi\rangle
\label{Xi}
\end{equation}
for $A, B \in \mathbb{R}$.
One can easily verify the following:
\begin{eqnarray}
&&\hspace{-7mm}
\Xi(A,B) \mbox{ is real},
\\
&&\hspace{-7mm}
\int \int_{-\infty}^{+\infty} \Xi(A,B) dAdB=1,
\label{sumAB}\\
&&\hspace{-7mm}
\int_{-\infty}^{+\infty} \hspace{-4mm}
\Xi(A,B) dB=
\Xi_{\hat A}(A),\
\int_{-\infty}^{+\infty}  \hspace{-4mm}
\Xi(A,B) dA
=\Xi_{\hat B}(B).
\label{sumAorB}\end{eqnarray}
In general, $\Xi(A,B)$ can take negative values.
If it is non-negative,
Eqs.~(\ref{sumAB}) and (\ref{sumAorB}) show that 
it can be interpreted as a 
coarse-grained joint probability density (cgJPD)
for $\hat{A}$ and $\hat{B}$.
In fact, as we will demonstrate in the following sections, 
$\Xi(A,B)$ becomes 
non-negative if
$W$ and $N$ are large enough,
for many states of interest.
Furthermore,
even if $W$ and $N$ are not large,
negative-valued regions of $\Xi(A,B)$ are small.
In this case, $\Xi(A,B)$ can be considered as
a coarse-grained {\it quasi} joint probability density (cgQJPD)
for $\hat A$ and $\hat B$.

The non-negativity of $\Xi(A,B)$ 
becomes obvious as $W \to \infty$, for which 
$\Xi(A,B) \sim w(A)w(B) \ge0$ for all $A,B$.
For smaller $W$, 
the smallest value of $W$ 
that makes $\Xi(A,B)$ non-negative
depends on $\hat A$, $\hat B$ and $| \psi \rangle$. 
Therefore, in general, the non-negativity should be checked a posteriori.

We can also introduce $\Xi$ for 
$m$ ($\ge3$) hermitian additive operators 
$\hat A_1, \hat A_2, \cdots, \hat A_m$ by
\begin{eqnarray}
&&
\hspace{-8mm}
\Xi(A_1, \cdots, A_m) 
\nonumber\\
&& 
\hspace{-5mm}
\equiv
\frac{1}{m!} \sum_{\pi} \langle\psi|
\overline{\mathcal{P}}_{\hat A_{\pi(1)}}(A_{\pi(1)}) 
\cdots 
\overline{\mathcal{P}}_{\hat A_{\pi(m)}}(A_{\pi(m)}) 
|\psi\rangle,
\label{P3}
\end{eqnarray}
where the sum is taken over all permutations $\pi$ 
of the numbers $1, 2, \cdots, m$.

If $| \psi \rangle$ 
has $p=2$,
one can visualize 
structure of the macroscopic superpositions contained
in $|\psi\rangle$
by plotting $\Xi(A_1, \cdots, A_m)$ versus $(A_1,\cdots,A_m)$,
if $\hat{A}_1, \cdots, \hat{A}_m$ are appropriately taken.
We call such a plot a {\em visualization} of 
superpositions of macroscopically distinct states in $| \psi \rangle$.
An efficient method of finding 
appropriate 
$\hat{A}_1,\cdots,\hat{A}_m$
will be explained in the next
subsection.

%---------------------------------------------------
\subsection{Efficient method of finding appropriate operators}
\label{finding}

In principle,  
one can take any
hermitian additive operators $\hat{A}_1, \cdots, \hat{A}_m$,
and plot 
$\Xi(A_1, \cdots, A_m)$.
In this paper, however, we are interested in 
states having $p=2$, which contain 
superpositions of macroscopically distinct states.
Such superpositions are characterized 
by macroscopic fluctuations of certain 
additive operators (see Sec.~\ref{p} and Refs.~\cite{SM,Morimae}).
Therefore, 
as will be demonstrated in the next section,
we can visualize 
such superpositions 
by including macroscopically fluctuating operator(s)
in $\hat{A}_1, \cdots, \hat{A}_m$ of $\Xi(A_1,\cdots,A_m)$.
In this subsection, we present an efficient 
method of finding a set $\mathcal{S}$  
of macroscopically fluctuating hermitian
additive operators.

For a given pure state $|\psi\rangle$,
we diagonalize the VCM to obtain its eigenvalues,
$e_1\ge e_2\ge\cdots\ge e_{3N}$,
and eigenvectors.
From the eigenvectors, we construct a
complete orthogonal system:
$\{\{\tilde{c}_{\alpha l}^{~1}\},
\{\tilde{c}_{\alpha l}^{~2}\}, 
\cdots,
\{\tilde{c}_{\alpha l}^{~3N}\}\}$ 
($\alpha=x,y,z$; $l=1,2,\cdots,N$).
Here, 
$\{\tilde{c}_{\alpha l}^{~i}\}\in\mathbb{C}^{3N}$ 
is an eigenvector of the VCM
corresponding to $e_i$.
We assume that 
each $\tilde{c}_{\alpha l}^{~i}$ is asymptotically independent of $N$,
and that we can normalize 
$\{\tilde{c}_{\alpha l}^{~i}\}$
as 
$\sum_{\alpha l}|\tilde{c}_{\alpha l}^{~i}|^2=N$.
By taking an appropriate limit of $\{ \tilde{c}_{\alpha l}^{~i} \}$
as described in Appendix \ref{app:limit}, 
we obtain a vector $\{ c_{\alpha l}^{i} \}$,
whose elements are independent of $N$.
From this vector, we 
construct the additive operator:
\begin{equation}
\hat{A}_i
\equiv\sum_{l=1}^N\sum_{\alpha=x,y,z}c_{\alpha l}^i
\hat{\sigma}_\alpha(l).
\label{A_i}
\end{equation}
As shown in Appendix \ref{app:Aifm}, 
$\hat{A}_i$ fluctuates macroscopically
if and only if $e_i=\mathcal{O}(N)$.

If $e_i=\mathcal{O}(N)$ and $\hat{A}_i$ is hermitian,
we let $\hat{A}_i$ be an element of $\mathcal{S}$. 
If $e_i=\mathcal{O}(N)$ and $\hat{A}_i$ is non-hermitian,
on the other hand,
we decompose $\hat{A}_i$ into the real and the imaginary parts:
$\hat{A}_i=\hat{A}_i^{\rm re}+i\hat{A}_i^{\rm im}$,
where 
$\hat{A}_i^{\rm re}\equiv(\hat{A}_i+\hat{A}_i^\dagger)/2$ and  
$\hat{A}_i^{\rm im}\equiv(\hat{A}_i-\hat{A}_i^\dagger)/2i$.
It is known that  
$\hat{A}_i^{\rm re}$ and/or $\hat{A}_i^{\rm im}$
fluctuate(s) macroscopically~(see Ref.~\cite{Morimae} and Appendix A).
We let such macroscopically fluctuating part(s)
be an element(s) of $\mathcal{S}$.
In this way, we obtain
a set of macroscopically fluctuating hermitian additive
operators, e.g., as
$%\begin{eqnarray}
\mathcal{S}=\{\hat{A}_1,
\hat{A}_2^{\rm im},\hat{A}_3,\hat{A}_4^{\rm re},\hat{A}_4^{\rm im},
\hat{A}_5\}$.
Several examples of $\mathcal{S}$ will be given in the next section.
Any macroscopically fluctuating additive operator 
includes at least one element of $\mathcal{S}$ as a component
in the sense explained in Appendix \ref{app:S=min}.

The number of the elements of $\mathcal{S}$ is $O(N^0)$,
because $e_i\ge 0$ and $\sum_{i=1}^{3N}e_i=\sum_{l=1}^N\sum_
{\alpha=x,y,z}V_{\alpha l, \alpha l}\le 3N$.
One can obtain $\mathcal {S}$ efficiently, because
one has only to diagonalize the VCM, which is
a $3N\times3N$ hermitian matrix.

By including an element(s) of $\mathcal{S}$ into 
$\hat{A}_1,\cdots,\hat{A}_m$ of $\Xi(A_1,\cdots,A_m)$,
one can visualize superpositions of macroscopically distinct states
in $|\psi\rangle$
by plotting $\Xi(A_1,\cdots,A_m)$. 

%-------------------------------------------------------------------------------
\section{Examples}\label{examples}
To demonstrate usefulness of the visualization method, 
we visualize 
four states having $p=2$ in this section.

%--------------------------------------------------------------
\subsection{XY model}
\label{XY}
First, we visualize the exact ground state of the XY model
on a two-dimensional square lattice of $N$ sites.
The Hamiltonian is 
\begin{eqnarray}
\hat{H}=-\sum_{<l,l'>}\left[\hat{\sigma}_x(l)\hat{\sigma}_x(l')+
\hat{\sigma}_y(l)\hat{\sigma}_y(l')\right],
\label{H_XY}
\end{eqnarray}
where $<l,l'>$ denotes the nearest neighbors.
If $N$ is finite, 
`ground states' obtained by the mean-field approximation 
are very different from 
the exact ground state~\cite{SMU(1),Koma,Oitmaa,Miyashita,Yukalov}.
These mean-field ground states are
degenerate  
symmetry-breaking states with non-zero order parameters.
They are separable states,
because the mean-field approximation neglects
the correlations between sites.
On the other hand,
the exact ground state is unique, symmetric,
and has $p=2$~\cite{SMU(1),Koma,Oitmaa,Miyashita,Morimae}. 
We visualize the exact ground state.

By numerical calculations, we find that $e_1=e_2=\mathcal{O}(N)$, 
$e_i=o(N)$ $(i\ge3)$,
$\hat{A}_1=\sum_{l=1}^N\hat{\sigma}_x(l)\equiv \hat{M}_x$,
and 
$\hat{A}_2=\sum_{l=1}^N\hat{\sigma}_y(l)\equiv \hat{M}_y$.
Hence,
\begin{eqnarray}
{\mathcal S}=\{\hat{M}_x,\hat{M}_y\}.
\label{XYA_0}
\end{eqnarray}

\begin{figure}[htbp]
\includegraphics[width=0.5\textwidth]{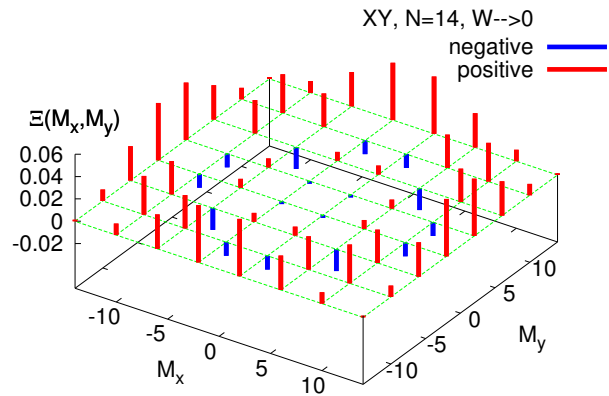}
\caption{(Color online) $\Xi(M_x,M_y)$ with $W\to0$ for
the exact ground state 
of the XY model on a two-dimensional square lattice with $N=14$.
$c\delta(0)$ [$c\in \mathbb{R}$] is represented by 
a vertical line with height $c$.
Positive values are represented by red vertical lines, whereas
negative values are represented by blue vertical lines.}
\label{XY14W0.eps}
\end{figure}

In Fig.~\ref{XY14W0.eps},
we plot $\Xi(M_x,M_y)$ for $N=14$ 
without coarse-graining, i.e., $W\to0$, 
for which the coarse-graining function of Eq.~(\ref{eq:w}) 
becomes the delta function $\delta(X)$.
In the figure, 
$c\delta(0)$ [$c\in \mathbb{R}$] is represented by 
a vertical line with height $c$.
Positive values are represented by red vertical lines, whereas
negative values are represented by blue vertical lines.
Because $\Xi(M_x,M_y)$ takes negative values at some points,
it is not a JPD for $\hat{M}_x$ and $\hat{M}_y$.

\begin{figure}[htbp]
\includegraphics[width=0.5\textwidth]{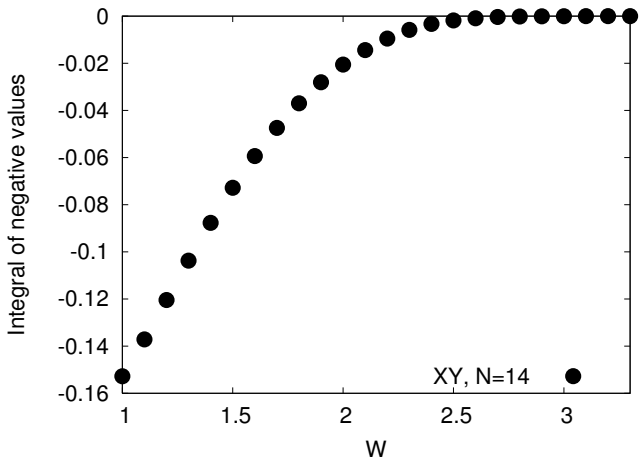}
\caption{The integral $I_{\hat{M_x},\hat{M}_y}$ 
of negative values of 
$\Xi(M_x,M_y)$ versus $W$ for
the exact ground state
of the XY model on a two-dimensional square lattice with $N=14$.}
\label{XYnegativeintegral.eps}
\end{figure}

However, negative values are
expected to approach 0 as $W$ is increased.
To see this, we plot
in Fig.~\ref{XYnegativeintegral.eps} 
the integral $I_{\hat{M}_x,\hat{M}_y}$
of negative values
versus $W$,
where $I_{\hat{A},\hat{B}}$ is defined by
\begin{eqnarray}
I_{\hat{A},\hat{B}}
\equiv\int\int_{-\infty}^{+\infty}dAdB\frac{\Xi(A,B)-|\Xi(A,B)|}{2}.
\label{integral}
\end{eqnarray}
It is seen that
$I_{\hat{M}_x,\hat{M}_y}$ indeed approaches 0 as $W$ is increased.
$\Xi(M_x,M_y)$ therefore becomes a cgJPD if $W$ is sufficiently large.

\begin{figure}[htbp]
\includegraphics[width=0.5\textwidth]{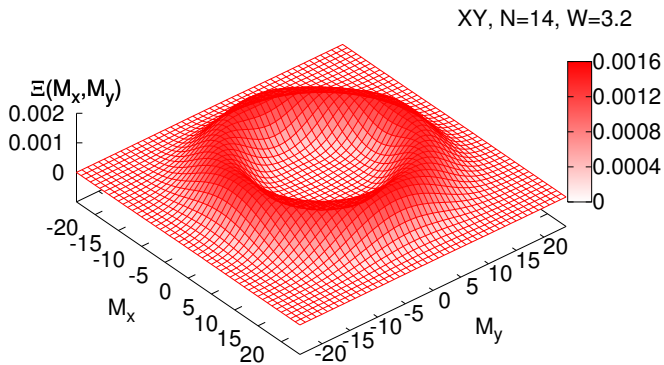}
\caption{(Color online) $\Xi(M_x,M_y)$ with $W=3.2$ for
the exact ground state 
of the XY model on a two-dimensional square lattice with $N=14$.}
\label{XY14W3_2.eps}
\end{figure}

For example, 
we plot $\Xi(M_x,M_y)$ with $W=3.2$
in Fig.~\ref{XY14W3_2.eps}.
In this case, $\Xi(M_x,M_y)$ is non-negative, and therefore 
it is a cgJPD. 
From this figure, we can clearly understand the
structure of the superposition of macroscopically distinct states:
Many macroscopically distinct states 
which have
macroscopically 
definite $\rm U(1)$ order parameters $(M_x,M_y)$
are so superposed 
that the ground state has the $\rm U(1)$ symmetry.

\begin{figure}[htbp]
\includegraphics[width=0.5\textwidth]{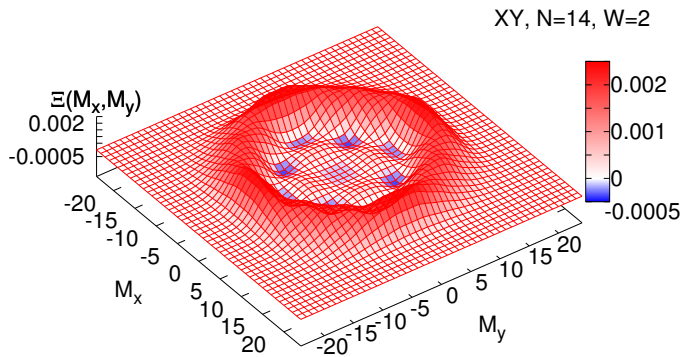}
\caption{(Color online) $\Xi(M_x,M_y)$ with $W=2$ for
the exact ground state 
of the XY model on a two-dimensional square lattice with $N=14$.
Positive-valued regions are colored red,
whereas negative-valued regions are colored 
blue.}
\label{XY14W2.eps}
\end{figure}

In Fig.~\ref{XY14W2.eps},
we plot $\Xi(M_x,M_y)$ with smaller $W$, $W=2$.
Because $\Xi(M_x,M_y)$ has negative-valued regions,
it is not a JPD.
However, since $|I_{\hat{M}_x,\hat{M}_y}|$ is small, 
$\Xi(M_x,M_y)$ is
well regarded as
a {\it quasi} JPD.
The figure shows the same $\rm U(1)$-symmetrical structure
as that of Fig.~\ref{XY14W3_2.eps}.

One can utilize either Fig.~\ref{XY14W3_2.eps} or Fig.~\ref{XY14W2.eps}
depending on the purpose:
When one wants a 
cgJPD, 
Fig.~\ref{XY14W3_2.eps} should be used.
On the other hand, 
when one wants to see more detailed structures, including 
quantum effects that make $\Xi$ negative, 
then Fig.~\ref{XY14W2.eps} (or, Fig.~\ref{XY14W0.eps})
would be better.
In this way, one can adjust $W$ to obtain 
a useful $\Xi$ according to the purpose.

%-------------------------------------------------------------
\subsection{Heisenberg antiferromagnet}
\label{HAF}
Second, we visualize the exact ground state  
of the Heisenberg antiferromagnet
on a two-dimensional square lattice of $N$ sites.
The Hamiltonian is 
\begin{eqnarray}
\hat{H}=\sum_{<l,l'>}\left[\hat{\sigma}_x(l)\hat{\sigma}_x(l')+
\hat{\sigma}_y(l)\hat{\sigma}_y(l')
+\hat{\sigma}_z(l)\hat{\sigma}_z(l')\right].
\label{H_HAF}
\end{eqnarray}
The `ground states' obtained by the mean-field 
approximation 
are degenerate, symmetry-breaking, and separable.
On the other hand, 
the exact ground state is unique, symmetric, 
and has $p=2$ if $N$ is finite~\cite{Koma,Morimae,Horsch,Marshall,Miyashita}.
We visualize the exact ground state.

By numerical calculations, we find that 
$e_1=e_2=e_3=\mathcal{O}(N)$,
$e_i=o(N)$ $(i\ge4)$,
$\hat{A}_1=\sum_{l=1}^N(-1)^l\hat{\sigma}_x(l)\equiv \hat{M}_x^{st}$,
$\hat{A}_2=\sum_{l=1}^N(-1)^l\hat{\sigma}_y(l)\equiv \hat{M}_y^{st}$,
and
$\hat{A}_3=\sum_{l=1}^N(-1)^l\hat{\sigma}_z(l)\equiv \hat{M}_z^{st}$.
Hence 
\begin{eqnarray}
{\mathcal S}
=\{\hat{M}_x^{st},\hat{M}_y^{st},\hat{M}_z^{st}\}.
\label{HAFA_0}
\end{eqnarray}
Because $\Xi(M_x^{st},M_y^{st},M_z^{st})$ is hard to plot,
we plot $\Xi(M_x^{st},M_y^{st})$.
Because of the rotational symmetry of the model, 
$\Xi(M_x^{st},M_y^{st})=\Xi(M_y^{st},M_z^{st})=\Xi(M_z^{st},M_x^{st})$.

\begin{figure}[htbp]
\includegraphics[width=0.5\textwidth]{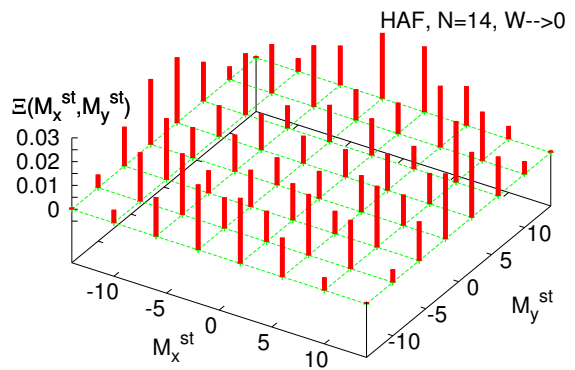}
\caption{(Color online) 
$\Xi(M_x^{st},M_y^{st})$ with $W\to0$ for
the exact ground state 
of the Heisenberg antiferromagnet on a 
two-dimensional square lattice with $N=14$.
$c\delta(0)$ [$c\in\mathbb{R}$] is represented by 
a vertical line with height $c$.}
\label{HAF14W0.eps}
\end{figure}

In Fig.~\ref{HAF14W0.eps},
we plot $\Xi(M_x^{st},M_y^{st})$  
with $W\to0$ for $N=14$.
Because 
$\Xi(M_x^{st},M_y^{st})$ 
is non-negative, 
it is a JPD
for $\hat{M}_x^{st}$ and $\hat{M}_y^{st}$.
It is also seen that many macroscopically distinct states
are superposed in the ground state.

\begin{figure}[htbp]
\includegraphics[width=0.5\textwidth]{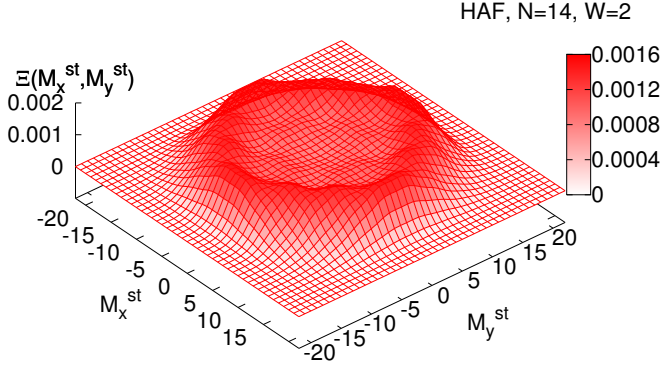}
\caption{(Color online)
$\Xi(M_x^{st},M_y^{st})$ with $W=2$ for
the exact ground state 
of the Heisenberg antiferromagnet on a 
two-dimensional square lattice with $N=14$.}
\label{HAF14W2.eps}
\end{figure}

By increasing $W$,
we obtain more understandable pictures.
For example,
in Fig.~\ref{HAF14W2.eps},
we plot $\Xi(M_x^{st},M_y^{st})$ with $W=2$.
It is seen that
many macroscopically distinct states are
so superposed that 
the ground state is symmetric,
like the ground state of the XY model.

%------------------------------------
\subsection{Shor's factoring algorithm}
\label{Shor}
Third, we visualize a state in Shor's factoring algorithm~\cite{Shor,Ekert}.
Let $I$ be an integer to be factored.
We use two quantum registers,
the first and the second registers,
which are composed of $N_1$ $(2\log_2I\le N_1 < 2\log_2I+1)$
and $N_2$ ($\log_2I\le N_2 < \log_2I+1$) qubits, respectively.
We denote the total number of qubits by $N=N_1+N_2$.
If the order $r$ is 6, for example, the state 
\begin{eqnarray}
\frac{1}{\sqrt{2^{N_1}}}
\sum_{a=0}^{2^{N_1}-1}|{\sf x}^a~{\rm mod}~I\rangle_2|a\rangle_1,
\label{ME}
\end{eqnarray}
which appears just after the modular exponentiation, 
has $p=2$~\cite{Ukena,Ukena2}.
Here, $| \cdots  \rangle_1$ and $| \cdots  \rangle_2$
represent the first and the second registers, respectively, 
and ${\sf x}\ ({\sf x}<I)$ is a randomly taken integer coprime to $I$.

For the states of $r=6$,
we numerically find that
$e_1=e_2=\mathcal{O}(N)$, 
$e_i=o(N)$ $(i\ge 3)$,
$\hat{A}_1=
\sqrt{\frac{3}{2}}\sum_{l=2}^{N_1}\hat{\sigma}_x(l)\equiv\hat{M}_x^{(1)}$, 
and 
$\hat{A}_2=
\sqrt{\frac{3}{2}}\sum_{l=2}^{N_1}(-1)^l\hat{\sigma}_y(l)
\equiv\hat{M}_y^{st (1)}$ (see Appendix B).
Here, the qubit states are $|0\rangle$ and $|1\rangle$
($\hat{\sigma}_z|0\rangle=-|0\rangle$, 
$\hat{\sigma}_z|1\rangle=|1\rangle$).
Hence 
\begin{eqnarray}
\mathcal {S}=\{\hat{M}_x^{(1)},\hat{M}^{st (1)}_y\}.
\label{ShorA_0}
\end{eqnarray} 
Because $\hat{M}_x^{(1)}$ and $\hat{M}_y^{st (1)}$ 
fluctuate macroscopically, $\hat{M}_x$ and $\hat{M}_y^{st}$
also fluctuate macroscopically~\cite{Ukena2}.
We here use $\hat{M}_x$ and $\hat{M}_y^{st}$ instead of 
$\hat{M}_x^{(1)}$ and $\hat{M}_y^{st (1)}$.

\begin{figure}[htbp]
\includegraphics[width=0.5\textwidth]{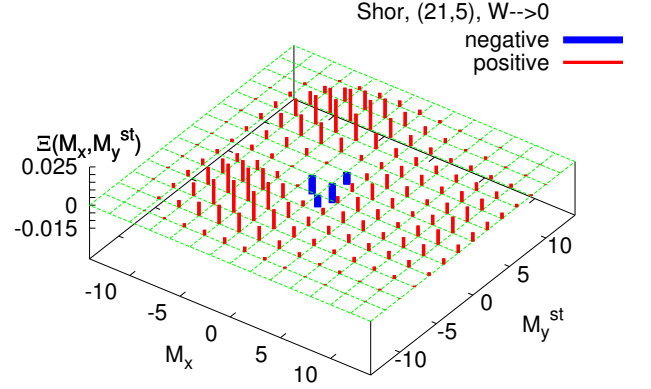}
\caption{(Color online)
$\Xi(M_x,M_y^{st})$ with $W\to0$ 
for
the state just after the modular exponentiation 
with $(I,{\sf x})=(21,5)$.
$c\delta(0)$ [$c\in\mathbb{R}$, $c>0$]
is represented by 
a vertical line with height $c$.
$c\delta(0)$ [$c\in\mathbb{R}$, $c<0$]
is represented by 
a vertical line with height $10c$,
in order to make negative values more visible.
Positive values are represented by red vertical lines, whereas
negative values are represented by blue vertical lines.}
\label{Shor21_5W0.eps}
\end{figure}

\begin{figure}[htbp]
\includegraphics[width=0.5\textwidth]{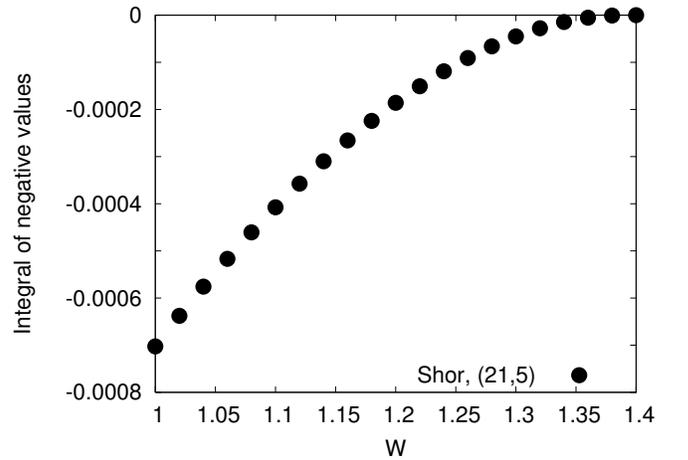}
\caption{The integral $I_{\hat{M}_x,\hat{M}_y^{st}}$
of negative values of
$\Xi(M_x,M_y^{st})$ versus $W$ 
for
the state just after the modular exponentiation 
with $(I,{\sf x})=(21,5)$.}
\label{Shornegativeintegral.eps}
\end{figure}

In Fig.~\ref{Shor21_5W0.eps},
we plot $\Xi(M_x,M_y^{st})$ 
with $W\to0$
for $(I,{\sf x})=(21,5)$.
Because $\Xi(M_x,M_y^{st})$ 
takes negative values at some points,
it is not a JPD.
To see the behavior of negative values, we
plot 
in Fig.~\ref{Shornegativeintegral.eps} 
the integral $I_{\hat{M}_x,\hat{M}_y^{st}}$
versus $W$.
We see again that $I_{\hat{M}_x,\hat{M}_y^{st}}$
approaches 0 as $W$ is increased.
$\Xi(M_x,M_y^{st})$ therefore becomes 
a cgJPD if $W$ is sufficiently large.

\begin{figure}[htbp]
\includegraphics[width=0.5\textwidth]{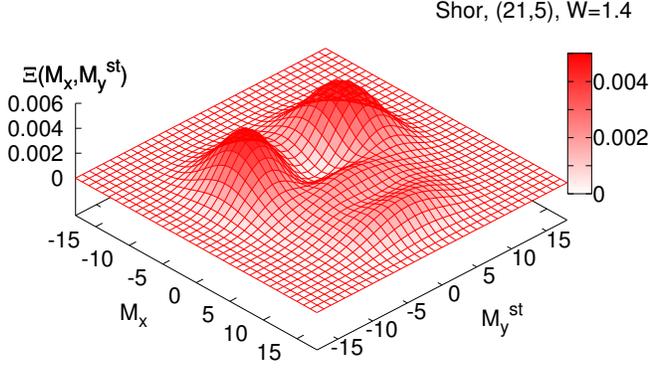}
\caption{(Color online)
$\Xi(M_x,M_y^{st})$ with $W=1.4$ for
the state just after the modular exponentiation with 
$(I,{\sf x})=(21,5)$.}
\label{Shor21_5W1_4.eps}
\end{figure}

In Fig.~\ref{Shor21_5W1_4.eps}, we plot 
$\Xi(M_x,M_y^{st})$ with $W=1.4$.
Because 
$\Xi(M_x,M_y^{st})$ is non-negative,
it is a cgJPD for $\hat{M}_x$ and $\hat{M}_y^{st}$.
There are four peaks, which represent a superposition of approximately four 
macroscopically distinct states.

\begin{figure}[htbp]
\includegraphics[width=0.5\textwidth]{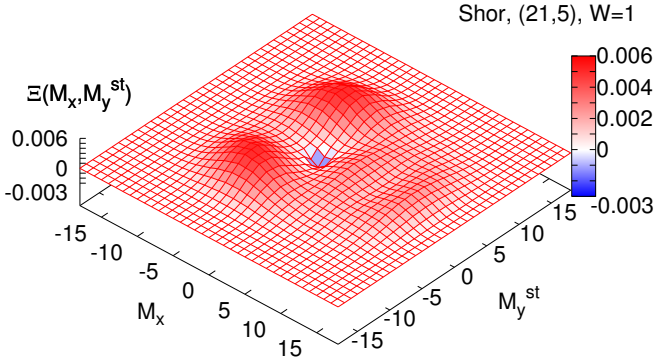}
\caption{(Color online)
$\Xi(M_x,M_y^{st})$ with $W=1$ for
the state just after the modular exponentiation with 
$(I,{\sf x})=(21,5)$.
Positive-valued regions are colored red, whereas
negative-valued regions are colored blue.
In order to make negative regions more visible,
negative values are multiplied by 10.}
\label{Shor21_5W1.eps}
\end{figure}

In Fig.~\ref{Shor21_5W1.eps}, we plot 
$\Xi(M_x,M_y^{st})$ with smaller $W$, $W=1$.
In this case, there is a negative-valued region.
However, because 
$|I_{\hat{M}_x,\hat{M}_y^{st}}|$
is small,
$\Xi(M_x,M_y^{st})$
is interpreted as a cgQJPD.
$\Xi(M_x,M_y^{st})$ again represents four peaks.
We have also observed
such a four-peak structure for some other values of $(I,{\sf x})$'s.

%--------------------------------------------------- 
\subsection{Grover's quantum search algorithm}
\label{Grover}
Finally, we visualize a state in 
Grover's quantum search algorithm~\cite{Grover,Nielsen}.
Let us consider the problem of finding a solution to the 
equation $f(x)=1$ among $2^N$ possibilities,
where $f(x)$ is a function, $f:\{0,1,\cdots,2^N-1\}\to\{0,1\}$.
These $2^N$ possibilities are indexed by 
$2^N$ computational basis states,
which are tensor products of 
$|0\rangle$ or $|1\rangle$ of
$N$ qubits.
Here,
$\hat{\sigma}_z|0\rangle=-|0\rangle$ 
and $\hat{\sigma}_z|1\rangle=|1\rangle$.
Let $|G_k\rangle$ be the state which appears 
after $k$ Grover 
iterations.
It was shown that 
if the number of the solutions is $\mathcal{O}(N^0)$, 
$|G_k\rangle$'s whose $k$ satisfies 
\begin{eqnarray}
\delta\le \frac{4k+2}{\sqrt{2^N}}\le \pi-\delta 
\label{k}
\end{eqnarray}
have $p=2$,
irrespective of which numbers are
the solutions~\cite{Ukena2}.
Here, $\delta$ is an arbitrary 
small positive constant being independent of $N$.

To be definite, 
we assume that the state 
$|1^{\otimes N}\rangle$
indexes the solution. 
Then
$|G_k\rangle$ is written as
\begin{eqnarray}
|G_k\rangle
&=&\cos\left(\frac{2k+1}{2}\theta\right)
\frac{1}{\sqrt{2^N-1}}
\left[|0^{\otimes N}\rangle+\cdots\right]\nonumber\\
&&~+\sin\left(\frac{2k+1}{2}\theta\right)|1^{\otimes N}\rangle,
\label{G_k}
\end{eqnarray}
where 
$\cos\frac{\theta}{2}=\sqrt{(2^N-1)/2^N}$, 
and 
$\left[|0^{\otimes N}\rangle+\cdots\right]$
is the equal-weight superposition of 
all computational basis states except for 
$|1^{\otimes N}\rangle$. 
Among many $k$'s which satisfy Eq.~(\ref{k}),
we use $k= R/2$ for even $R$,
and $k= R/2+0.5$ for odd $R$,
where $R\equiv {\rm CI}\left(\theta^{-1}\arccos\sqrt{2^{-N}}\right)$ 
is the number of total Grover iterations.
Here, ${\rm CI}(x)$ denotes the integer closest to the real number $x$.

We numerically find that 
$e_1=\mathcal{O}(N)$, 
$e_i=o(N)$ $(i\ge 2)$, 
and 
$\hat{A}_1=
\sum_{l=1}^N
\left[\frac{-1}{\sqrt{2}}\hat{\sigma}_x(l)
+\frac{1}{\sqrt{2}}\hat{\sigma}_z(l)\right]
\equiv\hat{M}_{x\mathchar`-z}$ (see Appendix B).
Hence 
\begin{eqnarray}
{\mathcal S}=\{\hat{M}_{x\mathchar`-z}\}.
\label{GroverA_0}
\end{eqnarray}
Because $\mathcal{S}$ has only one element, the macroscopic superposition
can be visualized by plotting 
the probability density 
$\langle G_k|{\mathcal P}_{\hat{M}_{x\mathchar`-z}}(M_{x\mathchar`-z})|G_k\rangle$.
However, because it is more interesting to plot $\Xi(M_{x\mathchar`-z},A)$,
where $\hat{A}$ is a hermitian additive operator, 
we plot $\Xi(M_{x\mathchar`-z},M_y)$ in this paper.
The shape of 
$\langle G_k|{\mathcal P}_{\hat{M}_{x\mathchar`-z}}(M_{x\mathchar`-z})|G_k\rangle$
can be deduced from that of 
$\Xi(M_{x\mathchar`-z},M_y)$.

\begin{figure}[htbp]
\includegraphics[width=0.5\textwidth]{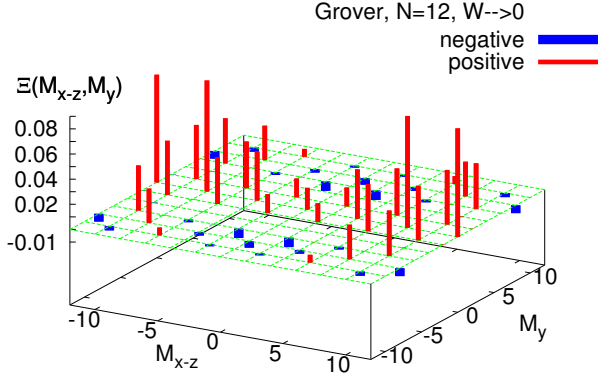}
\caption{(Color online)
$\Xi(M_{x\mathchar`-z},M_y)$ with $W\to0$ 
for
a state in Grover's quantum search algorithm with $N=12$.
$c\delta(0)$ [$c\in\mathbb{R}$] is represented by 
a vertical line with height $c$.
Positive values are represented by red vertical lines, whereas
negative values are represented by blue vertical lines.}
\label{Grover12W0.eps}
\end{figure}

\begin{figure}[htbp]
\includegraphics[width=0.5\textwidth]{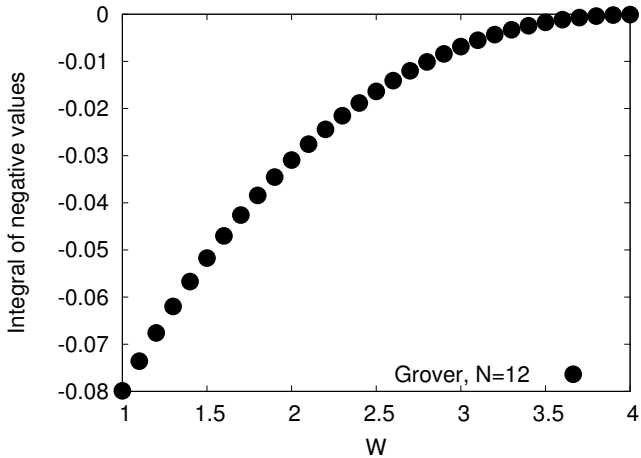}
\caption{ 
The integral $I_{\hat{M}_{x\mathchar`-z},\hat{M}_y}$ of negative values
of $\Xi(M_{x\mathchar`-z},M_y)$ versus $W$  
for
a state in Grover's quantum search algorithm with $N=12$.}
\label{Grover12negativeintegral.eps}
\end{figure}

In Fig.~\ref{Grover12W0.eps},
we plot $\Xi(M_{x\mathchar`-z},M_y)$ with $W\to0$ for $N=12$.
Because $\Xi$ takes negative values at some points,
it is not a JPD.
To see the behavior of the negative values, 
we plot 
in Fig.~\ref{Grover12negativeintegral.eps} 
the integral $I_{\hat{M}_{x\mathchar`-z},\hat{M}_y}$ 
versus $W$.
$I_{\hat{M}_{x\mathchar`-z},\hat{M}_y}$ approaches 0 as $W$ is increased.
$\Xi(M_{x\mathchar`-z},M_y)$ therefore becomes 
a cgJPD for $\hat{M}_{x\mathchar`-z}$ and $\hat{M}_y$
if $W$ is sufficiently large.

\begin{figure}[htbp]
\includegraphics[width=0.5\textwidth]{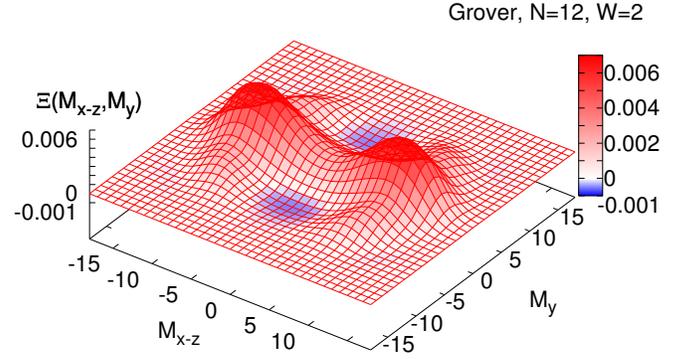}
\caption{
(Color online)
$\Xi(M_{x\mathchar`-z},M_y)$ with $W=2$ for
a state in Grover's quantum search algorithm with $N=12$.
Positive-valued regions are colored red, 
whereas negative-valued regions are
colored blue.}
\label{Grover12W2.eps}
\end{figure}

In Fig.~\ref{Grover12W2.eps},
we plot $\Xi(M_{x\mathchar`-z},M_y)$ with $W=2$.
Because there are small negative-valued regions,
it is a cgQJPD.
It is seen that the state is   
approximately a cat state, i.e., 
an equal-weight superposition of
two macroscopically distinct states. 
Although this information can also be obtained by plotting
$\langle G_k|\mathcal{P}_{\hat{M}_{x\mathchar`-z}}(M_{x\mathchar`-z})|G_k\rangle$,
we can see interesting structures of $|G_k\rangle$, including  
negative-valued regions,
by plotting $\Xi(M_{x\mathchar`-z},M_y)$. 

%---------------------------------------
\section{discussion}
\label{discussion}

\subsection{Non-negativity of $\Xi$}
In the previous section,
we have observed that 
an appropriate value of $W$ 
which makes $\Xi$ non-negative 
largely depends on the  
quantum state to be visualized.
For example,
$\Xi(M_x,M_y)$ for the ground state 
of the XY model
becomes non-negative with $W=3.2$,
whereas
$\Xi(M_x,M_y^{st})$ for the state in Shor's factoring algorithm 
becomes non-negative with smaller $W$,
$W=1.4$,
for the same value of $N$.
Furthermore, $\Xi(M_x^{st},M_y^{st})$ for
the ground state of the Heisenberg antiferromagnet
is non-negative with any $W$.
Therefore, 
in general,
one must find an appropriate value of $W$ a posteriori.

However, it is worth mentioning that 
a {\it sufficient} magnitude of $W$ which
makes $\Xi$ non-negative seems to be
$\mathcal{O}(N)$. 
To see this, consider the following three examples.

\begin{figure}[htbp]
\includegraphics[width=0.5\textwidth]{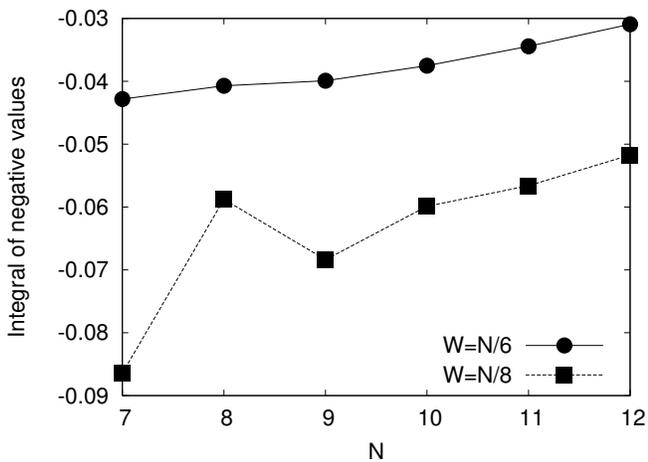}
\caption{ 
The integral $I_{\hat{M}_{x\mathchar`-z},\hat{M}_y}$ of negative values
of $\Xi(M_{x\mathchar`-z},M_y)$ versus $N$ with 
$W=N/6$ and $W=N/8$ 
for
states in Grover's quantum search algorithm.
Lines are guides to the eye.}
\label{GrovernegativeintegralN.eps}
\end{figure}

\begin{figure}[htbp]
\includegraphics[width=0.5\textwidth]{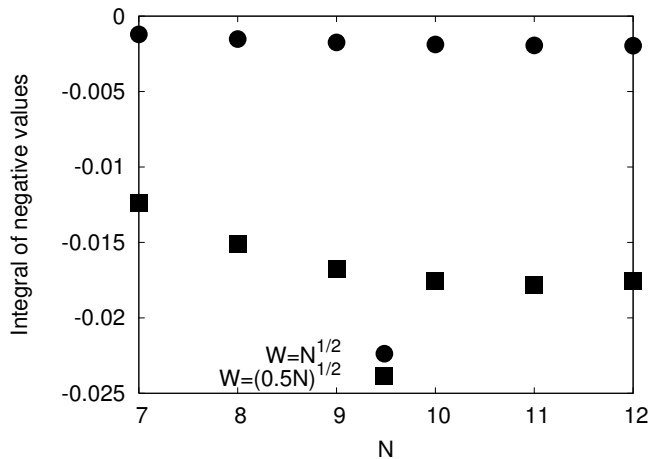}
\caption{ 
The integral $I_{\hat{M}_{x\mathchar`-z},\hat{M}_y}$ of negative values
of $\Xi(M_{x\mathchar`-z},M_y)$ versus $N$ with 
$W=\sqrt{N}$ and $W=\sqrt{0.5N}$
for
states in Grover's quantum search algorithm.}
\label{GrovernegativeintegralsqrtN.eps}
\end{figure}

{\it Example} 1: In Figs.~\ref{GrovernegativeintegralN.eps} 
and~\ref{GrovernegativeintegralsqrtN.eps},
we plot $I_{\hat{M}_{x\mathchar`-z},\hat{M}_y}$ versus $N$
for $|G_k\rangle$ of Eq.~(\ref{G_k})
with $W=\mathcal{O}(N)$ and $W=\mathcal{O}(\sqrt{N})$,
respectively.
Here, $k=R/2$ for even $R$, and $k=R/2+0.5$ for odd $R$.
It is seen that $I_{\hat{M}_{x\mathchar`-z},\hat{M}_y}$
approaches 0 as $N$ is increased if $W=\mathcal{O}(N)$,
whereas it does not approach 0 if $W=\mathcal{O}(\sqrt{N})$.

\begin{figure}[htbp]
\includegraphics[width=0.5\textwidth]{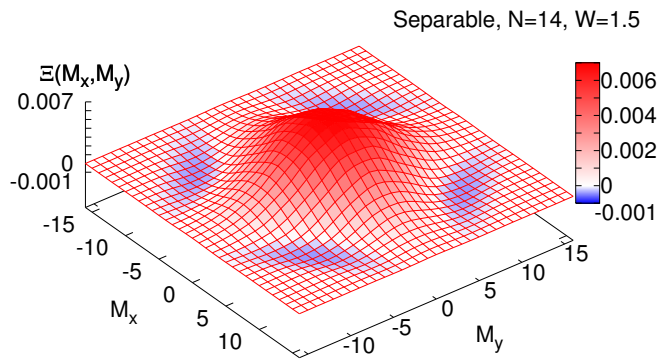}
\caption{ 
(Color online)
$\Xi(M_x,M_y)$ with $W=1.5$ for
the separable state $|0^{\otimes N}\rangle$
with $N=14$.
Positive-valued regions are colored red, whereas negative-valued regions
are colored blue.}
\label{sepa14W1_5.eps}
\end{figure}

\begin{figure}[htbp]
\includegraphics[width=0.5\textwidth]{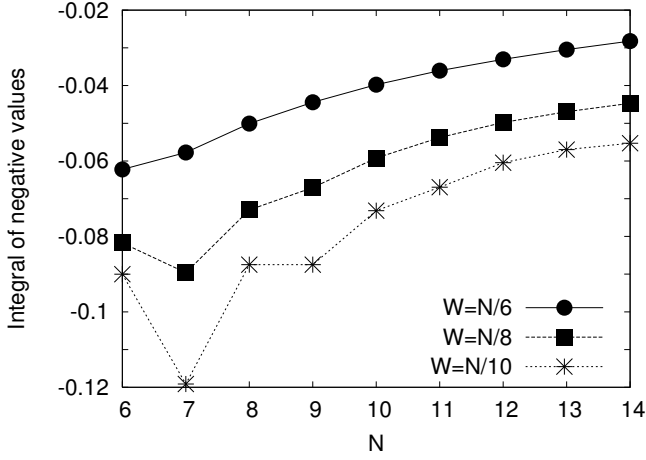}
\caption{ 
The integral $I_{\hat{M}_x,\hat{M}_y}$ of negative values
of $\Xi(M_x,M_y)$ versus $N$ with 
$W=N/6$, $W=N/8$, and $W=N/10$
for separable states $|0^{\otimes N}\rangle$.
Lines are guides to the eye.}
\label{sepanegativeintegralN.eps}
\end{figure}

\begin{figure}[htbp]
\includegraphics[width=0.5\textwidth]{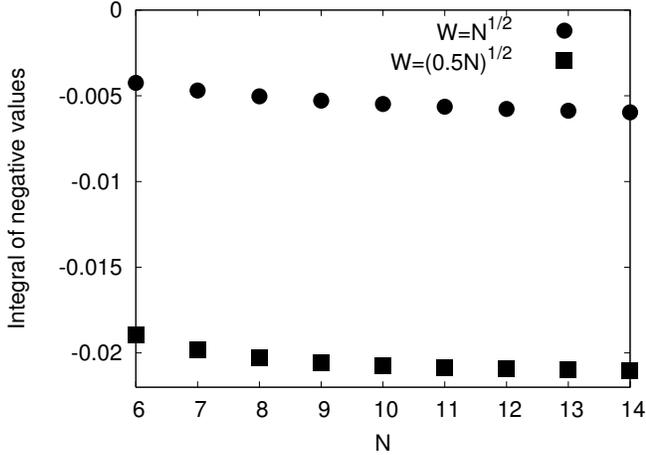}
\caption{ 
The integral $I_{\hat{M}_x,\hat{M}_y}$ of negative values
of $\Xi(M_x,M_y)$ versus $N$ with 
$W=\sqrt{N}$ and $W=\sqrt{0.5N}$
for separable states $|0^{\otimes N}\rangle$.}
\label{sepanegativeintegralsqrtN.eps}
\end{figure}

{\it Example} 2:
In Fig.~\ref{sepa14W1_5.eps}, we plot
$\Xi(M_x,M_y)$ 
for the separable state 
$|0^{\otimes N}\rangle$
with $W=1.5$ and $N=14$.
Here, $\hat{\sigma}_z|0\rangle=-|0\rangle$.
There are negative-valued regions.
In Figs.~\ref{sepanegativeintegralN.eps} 
and~\ref{sepanegativeintegralsqrtN.eps},
we plot $I_{\hat{M}_x,\hat{M}_y}$
versus $N$ 
with $W=\mathcal{O}(N)$ and $W=\mathcal{O}(\sqrt{N})$,
respectively.
We can see again that $I_{\hat{M}_x,\hat{M}_y}$
approaches 0 as $N$ is increased if $W=\mathcal{O}(N)$,
whereas it does not approach 0 if $W=\mathcal{O}(\sqrt{N})$.

\begin{figure}[htbp]
\includegraphics[width=0.5\textwidth]{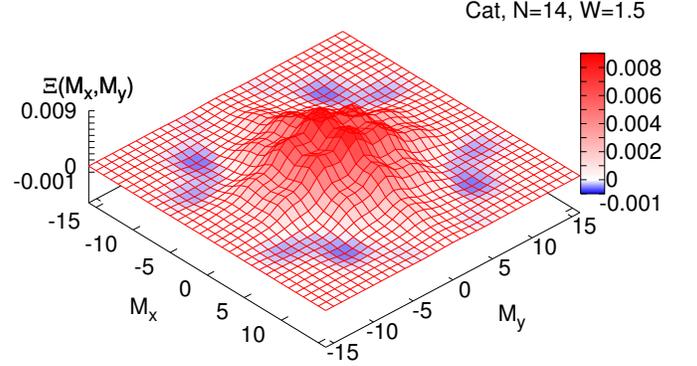}
\caption{ 
(Color online)
$\Xi(M_x,M_y)$ with $W=1.5$ for
the cat state 
$\frac{1}{\sqrt{2}}|0^{\otimes N}\rangle+
\frac{1}{\sqrt{2}}|1^{\otimes N}\rangle$
with $N=14$.
Positive-valued regions are colored red, 
whereas negative-valued regions are
colored blue.}
\label{cat14W1_5.eps}
\end{figure}

\begin{figure}[htbp]
\includegraphics[width=0.5\textwidth]{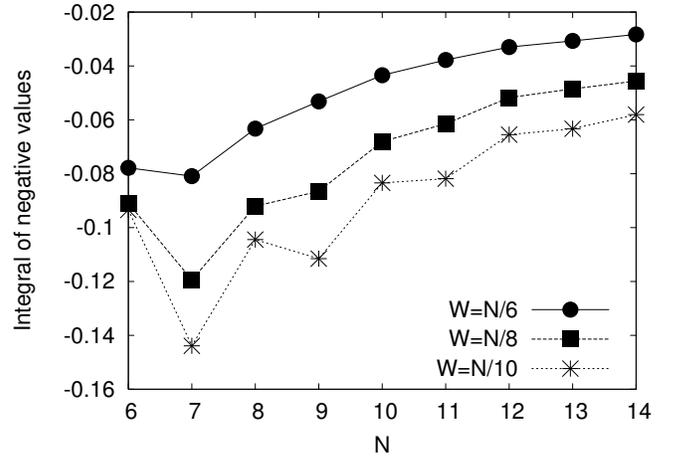}
\caption{ 
The integral $I_{\hat{M}_x,\hat{M}_y}$ of negative values
of $\Xi(M_x,M_y)$ versus $N$ with 
$W=N/6$, $W=N/8$, and $W=N/10$
for cat states 
$\frac{1}{\sqrt{2}}|0^{\otimes N}\rangle+
\frac{1}{\sqrt{2}}|1^{\otimes N}\rangle$.
Lines are guides to the eye.}
\label{catnegativeintegralN.eps}
\end{figure}

\begin{figure}[htbp]
\includegraphics[width=0.5\textwidth]{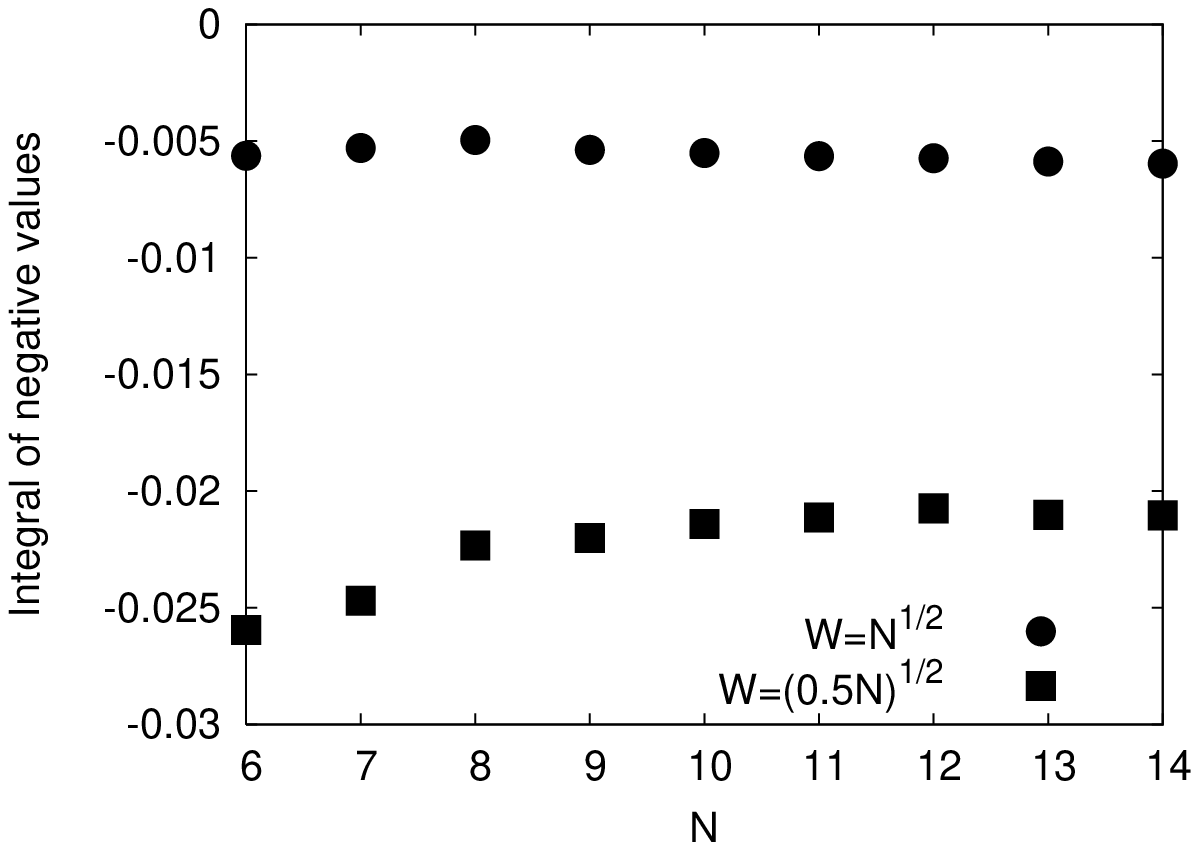}
\caption{ 
The integral $I_{\hat{M}_x,\hat{M}_y}$ of negative values
of $\Xi(M_x,M_y)$ versus $N$ with 
$W=\sqrt{N}$ and $W=\sqrt{0.5N}$
for cat states 
$\frac{1}{\sqrt{2}}|0^{\otimes N}\rangle+
\frac{1}{\sqrt{2}}|1^{\otimes N}\rangle$.}
\label{catnegativeintegralsqrtN.eps}
\end{figure}

{\it Example} 3: 
For the cat state
$\frac{1}{\sqrt{2}}|0^{\otimes N}\rangle
+\frac{1}{\sqrt{2}}|1^{\otimes N}\rangle$,
in which $\hat{M}_z$ fluctuates macroscopically,
$\Xi(M_z,M_x)$ and $\Xi(M_z,M_y)$ are non-negative. 
On the other hand,
$\Xi(M_x,M_y)$ can take negative values.
In Fig.~\ref{cat14W1_5.eps},
we plot $\Xi(M_x,M_y)$ 
with $W=1.5$ and $N=14$.
It is seen that there are negative-valued regions.
In Figs.~\ref{catnegativeintegralN.eps} 
and~\ref{catnegativeintegralsqrtN.eps},
we plot $I_{\hat{M}_x,\hat{M}_y}$  
with $W=\mathcal{O}(N)$ and $W=\mathcal{O}(\sqrt{N})$,
respectively.
$I_{\hat{M}_x,\hat{M}_y}$ again
approaches 0 as $N$ is increased if $W=\mathcal{O}(N)$,
whereas it does not approach 0 if $W=\mathcal{O}(\sqrt{N})$.

From these (and some other) examples,
it is expected that
$\mathcal{O}(N)$ is a sufficient 
magnitude of $W$ which makes $\Xi$ non-negative for sufficiently large $N$.
This expectation is reasonable,
because $W=\mathcal{O}(N)$ means that
the relative error of a measurement 
is independent of the system size $N$,
which is a usual situation for macroscopic systems.

Whether $\Xi(A_1,\cdots,A_m)$ is non-negative or not
depends also on which additive operators  
$\hat{A}_1,\cdots,\hat{A}_m$ are used.
For the ground state of the XY model,
for example, if we use $\hat{M}_x$ and $\hat{M}_z$
instead of $\hat{M}_x$ and $\hat{M}_y$,
$\Xi(M_x,M_z)$ 
is non-negative with any $W$,
because 
the ground state
is an eigenstate of $\hat{M}_z$ corresponding to the eigenvalue $M_z=0$,
hence 
$\Xi(M_x,M_z)=\Xi_{\hat{M}_x}(M_x)w(M_z)\ge 0$.

At the time of writing, however, we do not know a method of finding hermitian additive
operators and $W$ which make $\Xi$ non-negative for a given state.
To find such a method will be a subject of future studies.

%------------------------------------------------------
\subsection{Negative-valued regions of $\Xi$}
If $[\hat{A},\hat{B}]=0$,
$\Xi(A,B)$ is non-negative. 
It is therefore expected that 
negative-valued regions of $\Xi$ 
represent some quantum natures,
like those of the Wigner distribution function.

It seems that superposition of macroscopically distinct states studied in this paper
is not directly related to negative-valued regions.
For example, $\Xi(M_x^{st},M_y^{st})$ is non-negative
with any $W$
for the ground state of the Heisenberg antiferromagnet,
which has $p=2$.

In the previous subsection, on the other hand, we have seen that
$\Xi(M_x,M_y)$ has negative-valued regions for
the separable state $|0^{\otimes N}\rangle$. 
Because the separable state has no quantum nature other than the 
quantum coherence within each site,
the
negative-valued regions should represent
this quantum coherence.
This expectation is reasonable, because
$\Xi(M_x,M_y)$ 
is non-negative with any $W$ 
for the random state 
$\hat{\rho}_r\equiv\frac{1}{2^N}\hat{1}$,
which has neither entanglement nor quantum coherence. 
Here, we provisionally define $\Xi$
for a mixed state $\hat{\rho}$ by 
$
\Xi(A,B)\equiv
\frac{1}{2}{\rm Tr}\hat{\rho}
[\overline{\mathcal{P}}_{\hat A}(A) 
\overline{\mathcal{P}}_{\hat B}(B)
+
\overline{\mathcal{P}}_{\hat B}(B)
\overline{\mathcal{P}}_{\hat A}(A)] 
$.

Detailed analysis of negative-valued regions
is, however, beyond the scope of the
present paper.
It will also be a subject of future studies.

%-------------------------------------------------
\begin{acknowledgments}
This work was partially supported by Grant-in-Aid for
Scientific Research No.18-11581.
\end{acknowledgments}
%----------------------------------------------
\appendix
\section{$e_1=\mathcal{O}(N^{p-1})$}\label{app:e1}

In this appendix, we show that $e_1=\mathcal{O}(N^{p-1})$.
For a given pure state $|\psi\rangle$,
let $\{\tilde{c}_{\alpha l}^{~1}\}\in\mathbb{C}^{3N}$ 
be an eigenvector of the VCM
corresponding to the maximum eigenvalue $e_1$.
We normalize it as 
$\sum_{\alpha l}|\tilde{c}_{\alpha l}^{~1}|^2=N$.
From the eigenvector, we construct the operator:
\begin{eqnarray}
\hat{\sf A}_1\equiv
\sum_{\alpha l}\tilde{c}_{\alpha l}^{~1}\hat{\sigma}_{\alpha}(l).
\end{eqnarray}

If it is hermitian and all
$\tilde{c}_{\alpha l}^{~1}$'s are independent of $N$,
it gives the maximum of Eq.~(\ref{defp}).
Therefore, 
$\max_{\hat{A}}\langle\psi|(\Delta\hat{A})^2|\psi\rangle=e_1N$
in Eq.~(\ref{defp}).
Hence $e_1=\mathcal{O}(N^{p-1})$.

If $\hat{\sf A}_1$ is non-hermitian and all
$\tilde{c}_{\alpha l}^{~1}$'s are independent of $N$,
we decompose it as:
$\hat{\sf A}_1=\hat{A}_1^{\rm re}+i\hat{A}_1^{\rm im}$,
where 
$\hat{A}_1^{\rm re}\equiv(\hat{\sf A}_1+\hat{\sf A}_1^\dagger)/2$
and 
$\hat{A}_1^{\rm im}\equiv(\hat{\sf A}_1-\hat{\sf A}_1^\dagger)/2i$.
Then 
$\langle\psi|(\Delta\hat{A}_1^{\rm re})^2|\psi\rangle=\mathcal{O}(e_1N)$ 
or 
$\langle\psi|(\Delta\hat{A}_1^{\rm im})^2|\psi\rangle=\mathcal{O}(e_1N)$,
because
\begin{eqnarray}
\|\Delta\hat{\sf A}_1|\psi\rangle\|
&=&
\|\Delta\hat{A}_1^{\rm re}|\psi\rangle+
i\Delta\hat{A}_1^{\rm im}|\psi\rangle\|\nonumber\\
&\le&
\|\Delta\hat{A}_1^{\rm re}|\psi\rangle\|+
\|\Delta\hat{A}_1^{\rm im}|\psi\rangle\|.
\end{eqnarray}
Assume that 
$\langle\psi|(\Delta\hat{A}_1^{\rm re})^2|\psi\rangle=\mathcal{O}(e_1N)$. 
Because $\hat{\sf A}_1$ is additive,
$\hat{A}_1^{\rm re}$ is also additive. 
Then 
$\max_{\hat{A}}\langle\psi|(\Delta\hat{A})^2|\psi\rangle
=\mathcal{O}(e_1N)$ in Eq.~(\ref{defp}). 
Hence $e_1=\mathcal{O}(N^{p-1})$.

If some
$\tilde{c}_{\alpha l}^{~1}$'s depend on $N$,
we compose the additive operator
\begin{eqnarray}
\hat{A}_1\equiv
\sum_{\alpha l}c_{\alpha l}^{1}\hat{\sigma}_\alpha(l),
\end{eqnarray}
where 
$\{c_{\alpha l}^1\}$
is obtained by taking an appropriate limit of $\{\tilde{c}_{\alpha l}^{~1}\}$
as described in Appendix B. 
It can be shown that
$\langle\psi|\Delta\hat{A}_1^\dagger\Delta\hat{A}_1|\psi\rangle
=\mathcal{O}(e_1N)$.
In fact, 
by defining
$\tilde{d}_{\alpha l}^{~1}
\equiv \tilde{c}_{\alpha l}^{~1}-c_{\alpha l}^1$,
\begin{eqnarray}
\langle\psi|\Delta\hat{A}_1^\dagger\Delta\hat{A}_1|\psi\rangle
&=&\sum_{\alpha l \beta l'}
c_{\alpha l}^{1*}
c_{\beta l'}^{1}
V_{\alpha l,\beta l'}\nonumber\\
&=&\sum_{\alpha l \beta l'}
(\tilde{c}_{\alpha l}^{~1*}-\tilde{d}_{\alpha l}^{~1*})
(\tilde{c}_{\beta l'}^{~1}-\tilde{d}_{\beta l'}^{~1})
V_{\alpha l,\beta l'}\nonumber\\
&=&e_1N
+e_1o(N)
+\sum_{\alpha l\beta l'}
\tilde{d}_{\alpha l}^{~1*}\tilde{d}_{\beta l'}^{~1}
V_{\alpha l,\beta l'}\nonumber\\
&=&\mathcal{O}(e_1N),
\end{eqnarray}
where we have used the facts that
$\{\tilde{c}_{\alpha l}^{~1}\}$ is an eigenvector of 
the VCM
corresponding to $e_1$,
that 
$\sum_{\alpha l}|\tilde{c}_{\alpha l}^{~1}|^2=N$,
that 
$\tilde{d}_{\alpha l}^{~1}=o(N^0)$,
and that 
$0\le\sum_{\alpha l\beta l'}
\tilde{d}_{\alpha l}^{~1*}
\tilde{d}_{\beta l'}^{~1}
V_{\alpha l,\beta l'}\le e_1N$.
If $\hat{A}_1$ is hermitian,
$\max_{\hat{A}}\langle\psi|(\Delta\hat{A})^2|\psi\rangle
=\mathcal{O}(e_1N)$ in Eq.~(\ref{defp}).
Hence $e_1=\mathcal{O}(N^{p-1})$.
If $\hat{A}_1$ is not hermitian, its real or imaginary part
gives $\mathcal{O}(e_1N)$ fluctuation.
Therefore, 
$\max_{\hat{A}}\langle\psi|(\Delta\hat{A})^2|\psi\rangle
=\mathcal{O}(e_1N)$ in Eq.~(\ref{defp}).
Hence $e_1=\mathcal{O}(N^{p-1})$.

In conclusion, we have shown that $e_1=\mathcal{O}(N^{p-1})$.
%--------------------------------------------
\section{Composition of $\{c_{\alpha l}^i\}$ from $\{\tilde{c}_{\alpha l}^{~i}\}$}
\label{app:limit}

By diagonalizing the VCM,
one obtains %the $3N$-dimensional vector 
$\{ \tilde{c}_{\alpha l}^{~i} \}\in\mathbb{C}^{3N}$ 
corresponding to $e_i$.
Each element $\tilde{c}_{\alpha l}^{~i}$ generally depends on 
$N$, 
whereas
$c_{\alpha l}^{i}$'s composing 
the additive operator $\hat A_i$ through Eq.~(\ref{A_i}) should be 
independent of $N$.
We can deduce $c_{\alpha l}^{i}$ 
from $\tilde{c}_{\alpha l}^{~i}$
simply as follows.

Let us define a parameter $\nu \equiv l/N$, and 
denote $\tilde{c}_{\alpha l}^{~i}$
by $\tilde{c}_{\alpha \nu}^{~i}(N)$.
We take the following limit:
\begin{equation}
\lim_{N' \to \infty} \tilde{c}_{\alpha \nu}^{~i}(N')
\equiv 
c_{\alpha}^{i}(\nu),
\end{equation}
where  
$\nu$ is kept constant in this limit.
Then, $c_{\alpha l}^{i}$ is given by 
$c_{\alpha l}^{i}=c_{\alpha}^{i}(l/N)$.
Note that 
a small number [$=O(N^0)$] of 
elements among $3N$ elements of $\{ c_{\alpha l}^{i} \}$
can be modified, because
it does not alter the leading term (with respect to $N$) of
$\langle\psi|\Delta\hat{A}_i^\dagger\Delta\hat{A}_i|\psi\rangle$.
Using this property, 
we can adjust $\hat A_i$ for our convenience.

For the state of Eq.~(\ref{ME}) with $r=6$, for example, 
\begin{eqnarray}
\tilde{c}^{~1}_{\alpha l}=
\left\{
\begin{array}{cc}
\sqrt{N/(N_1-1)}
&(\alpha=x;\ 2 \leq l \leq N_1),\\
0&({\rm others}),
\end{array}
\right.
\label{eq:tc1}\\
\tilde{c}^{~2}_{\alpha l}=
\left\{
\begin{array}{cc}
(-1)^l\sqrt{N/(N_1-1)}
&(\alpha=y;\ 2 \leq l \leq N_1),\\
0&({\rm others}).
\end{array}
\right.
\label{eq:tc2}
\end{eqnarray}
We therefore obtain
\begin{eqnarray}
c^1_{\alpha l}=
\left\{
\begin{array}{cc}
\sqrt{3/2}
&(\alpha=x;\ 1 \leq l \leq N_1),\\
0&({\rm others}),
\end{array}
\right.
\\
c^2_{\alpha l}=
\left\{
\begin{array}{cc}
(-1)^l\sqrt{3/2}
&(\alpha=y;\ 1 \leq l \leq N_1),\\
0&({\rm others}).
\end{array}
\right.
\end{eqnarray}
Or, we can modify the $l=1$ terms of these results as
$c^1_{\alpha 1}=\ c^2_{\alpha 1}=0$ in accordance with 
the $l=1$ terms of Eqs.~(\ref{eq:tc1}) and (\ref{eq:tc2}).
We have employed the latter forms in Sec.~\ref{Shor}.

Moreover,
for $|G_k\rangle$ with $k=R/2$ (even $R$) or $k=R/2+0.5$ (odd $R$),
\begin{eqnarray}
\tilde{c}^{~1}_{\alpha l}=
\left\{
\begin{array}{cc}
-a/\sqrt{a^2+b^2+c^2}&(\alpha=x;~1\le l\le N),\\
ib/\sqrt{a^2+b^2+c^2}&(\alpha=y;~1\le l\le N),\\
c/\sqrt{a^2+b^2+c^2}&(\alpha=z;~1\le l\le N).
\end{array}
\right.
\end{eqnarray}
Here, $a$, $b$ and $c$ are real numbers which depend on $N$. 
It is numerically shown that $\lim_{N\to\infty}(a-c)=0$ and 
$\lim_{N\to\infty}b/\sqrt{a^2+b^2+c^2}=0$.
We therefore obtain
\begin{eqnarray}
c^1_{\alpha l}=
\left\{
\begin{array}{cc}
-1/\sqrt{2}&(\alpha=x;~1\le l\le N),\\
0&(\alpha=y;~1\le l\le N),\\
1/\sqrt{2}&(\alpha=z;~1\le l\le N),
\end{array}
\right.
\end{eqnarray}
which has been used in Sec.~\ref{Grover}.

%-------------------------------------------
\section{
$\hat{A}_i$ 
fluctuates macroscopically if and only if
$e_i=\mathcal{O}(N)$}
\label{app:Aifm}

In this appendix,
we show that
$\hat{A}_i$ of Eq.~(\ref{A_i})
fluctuates macroscopically if and only if
$e_i=\mathcal{O}(N)$.
Using $c_{\alpha l}^i$ of Appendix \ref{app:limit}, 
we define $\tilde{d}_{\alpha l}^{~i}
\equiv \tilde{c}_{\alpha l}^{~i}-c_{\alpha l}^i$.
Then
\begin{eqnarray}
\langle\psi|\Delta\hat{A}_i^\dagger \Delta\hat{A}_i|\psi\rangle
&=&\sum_{\alpha l \beta l'}
c_{\alpha l}^{i *}
c_{\beta l'}^{i}
V_{\alpha l,\beta l'}\nonumber\\
&=&\sum_{\alpha l \beta l'}
(\tilde{c}_{\alpha l}^{~i*}-\tilde{d}_{\alpha l}^{~i*})
(\tilde{c}_{\beta l'}^{~i}-\tilde{d}_{\beta l'}^{~i})
V_{\alpha l,\beta l'}\nonumber\\
&=&e_iN
+e_io(N)+o(N^2),
\end{eqnarray}
where we have used the facts that
$\{\tilde{c}_{\alpha l}^{~i}\}$ 
is an eigenvector of the VCM
corresponding to $e_i$,
that $\sum_{\alpha l}|\tilde{c}_{\alpha l}^{~i}|^2=N$,
and that $\tilde{d}_{\alpha l}^{~i}=o(N^0)$.
Therefore, if $e_i=\mathcal{O}(N)$ then 
$\langle\psi|\Delta\hat{A}_i^\dagger \Delta\hat{A}_i|\psi\rangle
=\mathcal{O}(N^2)$.
On the other hand, if $e_i=o(N)$ then 
$\langle\psi|\Delta\hat{A}_i^\dagger \Delta\hat{A}_i|\psi\rangle
=o(N^2)$.
%--------------------------------------------
\section{
any macroscopically fluctuating additive operator includes an element of $\mathcal{S}$
}\label{app:S=min}

For an additive operator 
$\hat{A}=\sum_{\alpha l}c_{\alpha l}\hat{\sigma}_\alpha(l)$,
the coefficient 
vector $\{c_{\alpha l}\} \in\mathbb{C}^{3N}$ can be expressed as
a linear combination of $\{\tilde{c}_{\alpha l}^{~i}\}$'s:
$c_{\alpha l}=\sum_{i=1}^{3N}\xi_i\tilde{c}_{\alpha l}^{~i}$,
where $\xi_i$'s are coefficients satisfying 
$\sum_i |\xi_i|^2=\mathcal{O}(N^0)$.
Assume that $\xi_i=o(N^0)$ if $e_i=\mathcal{O}(N)$ ($i=1,\cdots,3N$). Then
\begin{equation}
\langle\psi|\Delta\hat{A}^\dagger\Delta\hat{A}|\psi\rangle
=N\sum_i|\xi_i|^2e_i=o(N^2),\label{apC}
\end{equation}
where 
we have used the facts that 
$\{\tilde{c}_{\alpha l}^{~i}\}$'s are orthogonal eigenvectors of the VCM,
and that $\sum_{\alpha l}|\tilde{c}_{\alpha l}^{~i}|^2=N$.
Equation (\ref{apC}) shows that $\hat{A}$
does not fluctuate macroscopically.

In other words, 
if $\hat{A}$ fluctuates macroscopically,
its coefficient vector $\{c_{\alpha l}\}$ includes
at least one 
$\{\tilde{c}_{\alpha l}^{~i}\}$ 
whose $e_i=\mathcal{O}(N)$
as a component
of the linear combination with the weight $\xi_i=\mathcal{O}(N^0)$.
Therefore,
\begin{eqnarray}
\hat{A}
&=&\sum_{\alpha l}
[\cdots+
\xi_i\tilde{c}^{~i}_{\alpha l}\hat{\sigma}_\alpha(l)
+\cdots]\nonumber\\
&=&\sum_{\alpha l}[
\cdots+
\xi_i(c^i_{\alpha l}+\tilde{d}^{~i}_{\alpha l})\hat{\sigma}_\alpha(l)
+\cdots]\nonumber\\
&=&
\cdots+
\xi_i\hat{A}_i
+\cdots,
\end{eqnarray}
which shows that 
$\hat{A}$ includes $\hat{A}_i$ (hence also $\hat{A}_i^{\rm re}$ and $\hat{A}_i^{\rm im}$)
with the weight $\xi_i=\mathcal{O}(N^0)$.
In this sense, 
at least one element of $\mathcal{S}$ is `included' in $\hat{A}$,
if $\hat{A}$ fluctuates macroscopically.

%---------------------------------------------


\begin{thebibliography}{99}
\bibitem{O}
In this paper, we use three symbols 
$\mathcal{O}$, $O$, and $o$
to represent 
asymptotic behavior of a function $f(N)$
as $N \to \infty$:
$f(N)=\mathcal{O}(N^n)$ if $\lim_{N\to\infty}f(N)/N^n=$ constant $\neq 0$, 
$f(N)=O(N^n)$ if $\lim_{N\to\infty}f(N)/N^n$ is finite, 
and 
$f(N) = o(N^n)$ if $\lim_{N\to\infty}f(N)/N^n=0$.
\bibitem{xp}
This point can be seen by the following trivial example.
Dividing 
$[ \hat x, \hat p] = i \hbar$ 
by a large number $N$ yields
$
\left\| \left[ \hat{x}/N, \hat{p}/N \right] \right\|
= \hbar/N^2.
$
However,  
the noncommutativity of $\hat x$ and $\hat p$
{\em can} be detected by high-resolution experiments.
\end{thebibliography}
\end{document}